\documentstyle[prl,aps,amssymb,psfig]{revtex}

\makeatletter
\def\setcaption#1{\def\@captype{#1}}
\makeatother

\newcommand{\SuperK}    {Super--Kamiokande }

\newcommand{\pknu}      {$p \to \bar{\nu} K^+$ }
\newcommand{\munu}      {$K^+ \to \mu^+ \nu_{\mu}$ }
\newcommand{\pipi}      {$K^+ \to \pi^+ \pi^0$ }

\begin{document}
\title{Search for proton decay through \pknu  in 
        a large water Cherenkov detector}
\draft
\date{\today}
\maketitle
{\center \large The Super-Kamiokande Collaboration\\}

\begin{center}
\newcounter{foots}
Y.Hayato$^u$, M.Earl$^b$,
Y.Fukuda$^a$, T.Hayakawa$^a$, K.Inoue$^a$, 
K.Ishihara$^a$, H.Ishino$^a$, Y.Itow$^a$,
T.Kajita$^a$, J.Kameda$^a$
% ...
\addtocounter{foots}{1}
\addtocounter{foots}{1}
\addtocounter{foots}{1} 
L.L.Wai$^{w,\fnsymbol{foots}}$, 
R.J.Wilkes$^w$, K.K.Young$^{w,\dagger}$

\footnotesize \it

$^a$Institute for Cosmic Ray Research, University of Tokyo, Tanashi,
Tokyo 188-8502, Japan\\
$^b$Department of Physics, Boston University, Boston, MA 02215, USA  \\
% ...
$^u$Department of Physics, Tokyo Institute of Technology, Meguro, 
Tokyo 152-8551, Japan \\
% ...
$^w$Department of Physics, University of Washington,    
Seattle, WA 98195-1560, USA    \\
\end{center}

\begin{abstract}
	We present results of a search for
proton decays, $p \rightarrow \bar{\nu} K^+$, using data from 
a 33 kton$\cdot$year exposure of the Super--Kamiokande detector.
Two decay modes of the kaon, $K^+ \rightarrow \mu^+ \nu_{\mu}$ and
$K^+ \rightarrow \pi^+ \pi^0$, were studied.  The data were
consistent with the background expected from atmospheric neutrinos; 
therefore a lower limit on the
partial lifetime of the proton $\tau / B(p  \rightarrow \bar{\nu} K^+)$
was found to be $6.7 \times 10^{32}$ years at 90\% confidence level.
\end{abstract}

\par 
One of the most unique predictions of Grand Unified Theories (GUTs)
is baryon number violation. 
The minimal SU(5) GUT\cite{SU5GUT} predicts the dominant decay mode
to be $p \rightarrow e^+ \pi^0$
with a predicted lifetime shorter than $\sim 10^{31}$ years.
The current experimental lower limit is about 100 times longer than
this\cite{SKEPI}.
Furthermore, the weak mixing angle predicted by this model does not
agree with the experimental value and the three running coupling 
constants of the strong and electroweak forces do not meet exactly at a single 
point\cite{SIN2TWLEP}.
Alternatively, the minimal supersymmetric (SUSY) SU(5) GUT 
makes a prediction for the weak mixing angle which is much closer 
to experimental results and predicts the lifetime of 
the proton decay into $e^+ \pi^0$ to be more than four orders
of magnitude longer than that in non-SUSY minimal SU(5) GUT\cite{SUSYSU5}. 
The minimal SUSY SU(5) model predicts the 
proton decay mode $p \rightarrow \bar{\nu} K^+$  to be dominant 
with the partial lifetime prediction varying from  $\mathcal{O}$$(10^{29})$ to 
$\mathcal{O}$$(10^{35})$ yr\cite{SUSYSU5}.
\par
In this letter, we report the result of the search for proton decay 
through the channel $p \rightarrow  \bar{\nu} K^+$ in 535 live-days 
of data in Super--Kamiokande, corresponding to a 33 kt$\cdot$year exposure.

\par
The \SuperK  detector is a ring imaging water Cherenkov detector 
located in Kamioka Observatory, ICRR, Univ. of Tokyo,
\mbox{1000 m} (\mbox{2700 m} water equivalent) below the peak of Mt. Ikenoyama 
near Kamioka, Japan.  
50 kilotons of ultra pure water are held within a stainless steel
tank of height \mbox{41.4 m} and diameter \mbox{39.3 m}.
The tank is optically separated into
two regions, the inner and outer detectors, by a photomultiplier tube (PMT) 
support structure and a pair of opaque plastic sheets.
The inner detector has a height of \mbox{36.2 m} and a diameter of 
\mbox{33.8 m}.  The
outer detector completely surrounds the inner detector and is used to identify 
incoming and outgoing particles.  On the wall of the inner detector, there are 
11146 \mbox{50 cm} inward facing PMTs which cover 40\% of the surface.
The outer detector is lined with 1885 \mbox{20 cm} outward 
facing PMTs equipped with \mbox{60 cm $\times$ 60 cm}  wavelength-shifter
plates to increase the collection efficiency of Cherenkov photons.

\par
 In the search for rare proton decays, neutrinos produced by
cosmic ray interactions in the upper atmosphere (atmospheric neutrinos)
typically represent the limiting background. Super-Kamiokande detects 
about 8 fully-contained atmospheric neutrino events per day\cite{atmpd_paper}.
These events have vertices which were reconstructed inside the 
22.5 kiloton fiducial volume of the detector, defined to be 2 m from the PMT
support structure.  All the events were required to have no coincident light 
in the outer detector.

 After the event selection, events were reconstructed using PMT pulse 
height and timing information to determine
the vertex position, 
number of Cherenkov rings, particle type,
momentum, and 
number of decay electrons.  A particle is classified as a 
showering($e$-like)
or a non-showering($\mu$-like) type\cite{atmpd_paper}.
Details of the detector, event selection, 
and event reconstruction are described in \cite{SKEPI} and \cite{atmpd_paper}.

\par
The absolute scale of the momentum reconstruction was checked with several
calibration sources such as the electrons from a LINAC\cite{LINAC}, 
decay electrons from
cosmic-ray muons which stopped in the detector volume (stopping muons), 
$dE/dx$ of stopping muons, 
and the reconstructed mass of $\pi^0$s produced
by atmospheric neutrino interactions. The error in
the absolute energy scale was estimated to be less than $\pm2.5\%$.
The time variation of the energy scale was checked with muon 
decay electrons and
was found to vary by $\pm1\%$ over the exposure period.

\par
The momentum of the $K^+$ from \pknu  is 340 MeV/$c$
and is below the threshold momentum for producing 
Cherenkov light in water.
Candidate events for this decay mode are therefore identified through the 
decay products of the $K^+$.  
 Due to the smallness of the hadronic cross-section of low-momentum $K^+$, 
we calculated that 97\% of the $K^+$ exit from the $^{16}$O nucleus without
interaction, and that 90\% of these decay at rest. In this paper,
we therefore searched for $K^+$ decays at rest through two dominant
modes: 
$K^+ \rightarrow \mu^+ \nu_\mu$ and $K^+ \rightarrow \pi^+ \pi^0$.
Two separate methods were used to search for 
$K^+ \rightarrow \mu^+ \nu_\mu$.

\par
We now describe the first method to search for \pknu ; \munu.
The momentum of the $\mu^+$ from the decay of the stopped $K^+$ is 
236 MeV/$c$. 
If a proton in the $p_{3/2}$ state of $^{16}$O decays, the 
remaining $^{15}$N nucleus is left in an excited state.  This state 
quickly decays, emitting a prompt 6.3 MeV $\gamma$-ray. The signal from the 
decay particles of the $K^+$ should be delayed relative to that of the 
$\gamma$-ray due to the lifetime of the $K^+$ \mbox{($\tau=12$ ns)}.
The probability of a 6.3 MeV $\gamma$-ray emission 
in a proton decay in oxygen is estimated 
to be 41\%\cite{GAMEMIT}. By requiring this prompt $\gamma$-ray, almost
all of the background events are eliminated.

\par
To determine the signal region and to estimate the detection efficiency, a
total of 2000 \pknu ; \munu Monte Carlo (MC) events were generated.  The same
reduction and reconstruction as for the data were applied to these events.  
For these proton decay MC events, 96\% were 
identified as having one ring. The resolution of the vertex fitting was 
52 cm and the probability of misidentification of the
particle type was 2.9\%.
Based on this simulation, a 
signal region of the reconstructed momentum of the $\mu^+$ was defined to 
be between 215 MeV/$c$ and 260 MeV/$c$.
Candidate events for this analysis were then selected by the 
following criteria:
{A1) one\ ring\, A2) $\mu$-like\, A3) with\ one\ decay\ electron,
A4) 215 MeV/$c$ $<$ momentum($\mu$) $<$ 260 MeV/$c$,
A5) Goodness of fit requirement,
and A6) detection of a prompt $\gamma$-ray. 
The A5 criterion required
a successful vertex fitting for the correct identification of the
prompt $\gamma$-ray.
The prompt $\gamma$-ray signal was selected using the number of hit PMTs
within a 12ns timing window which slides
between 12ns and 120ns before the $\mu$ signal.  Criterion A6 required 
the number of hit PMTs to be more than 7.
The total detection efficiency, including the selection criteria A1-A6,
the branching ratio of \munu (63.5\%), the
ratio of bound protons in oxygen to all of the protons in H$_2$O (80\%),
and the emission probability of the 6.3 MeV $\gamma$-ray (41\%), was estimated
to be 4.4\%.  
The reason for the relatively low efficiency was due to the inefficiency
of the detection of the prompt $\gamma$-rays.
The number of background events was estimated using a
225 kt$\cdot$year equivalent sample of atmospheric neutrino MC events.
When estimating this background, the flux was normalized based on the
observed deficit of $\nu_\mu$'s.  The normalization
factors were 0.74 for $\nu_\mu$ charged current events and 1.17 for any other
neutrino induced events \cite{atmpd_paper}.
The background contamination was estimated to be 0.4 events/33 kt$\cdot$yr.
When the same criteria were applied to the real data, no events were observed.
Figure \ref{ivpekrk}
shows the number of hit PMTs within the timing region of the prompt 
$\gamma$-ray for proton decay MC events, simulated atmospheric neutrino 
events, and real data.

The partial lifetime ($\tau/B$) was calculated:
\begin{equation}
  \tau/\beta \ge (\Lambda\times \epsilon B_m) / N_{cand},\label{lifelim}
\end{equation}
where $\Lambda$ is the exposure in proton$\cdot$years,
$\epsilon B_m$ is the detection efficiency (4.4\%)
and $N_{cand}$ is the 90\% Poisson upper limit on the number of 
candidate events (2.3).  The lower limit of the partial lifetime 
obtained for this mode was found to be $2.1 \times 10^{32}$ yr at 90\% C.L.

\par
The overall detection efficiency of the prompt $\gamma$-ray tagging method is
rather small. Therefore, as a second method we searched for an 
excess of events of mono-energetic 236 MeV/c $\mu^+$'s produced by the
stopped $K^+$ decay for events with no observed prompt $\gamma$-ray.
For this analysis, the selection criteria were: A1 to A4 described
above and A7 no prompt gamma-ray signal.
The detection efficiency of this mode (including the branching ratio
of \munu) was estimated to be 40\%.

\par
To estimate the excess of proton-decay signal, the number of events 
in three momentum regions, 200 MeV/$c$ to 215 MeV/$c$, 215 MeV/$c$ to 
260 MeV/$c$, and 260 MeV/$c$ to 300 MeV/$c$, were summed 
separately for \pknu;\munu MC, atmospheric $\nu$ MC, and data.  
The $\chi^2$ method was applied to fit parameters. The $\chi^2$ function 
was defined as follows:
\begin{equation}
  \chi^2(a,b)=\sum_{i=1}^{3}
  \frac{[N^{data}_i-(a \cdot N^{atm\nu}_i + b \cdot N^{pdcy}_i)]^2}
        {N^{data}_i}
\end{equation}
where \mbox{$a$ and $b$} are the fit parameters, $N^{data}_i,N^{pdcy}_i,N^{atm\nu}_i$
are the numbers of events of real data, proton decay MC, and
atmospheric $\nu$ MC, respectively, in each momentum region $i$.

\par
The minimum $\chi^2$, $\chi^2_{min}=0.3$, was in the unphysical
region ($b\cdot N^{pdcy}_2 = -16.3$). 
The $\chi^2_{min}$ in the physical
region ($b\cdot N^{pdcy}_2 = 0$), $\chi^2_{min}(phys)$ was 2.0.
The data were therefore consistent with no excess of \pknu events. 
The 90\% C.L. upper limit on the number of proton decay
events($N_{cand}$) was obtained by requiring 
$\chi^2-\chi^2_{min}(phys)=3.7$ based on the prescription described in 
Ref. \cite{atmpd_osc}.
The 90\% upper limit on the number of candidates was estimated to be 13.3.
The momentum distribution of the events which satisfy criteria A1 to A3
is shown in Figure \ref{mom_dist_fitresult}. Also shown is the expected
signal of proton decay at the 90\% C.L.
The lower limit of the partial lifetime of
$p \rightarrow \bar{\nu}K^+; K^+\rightarrow\mu^+ \nu_\mu$ 
using the above method
was found to be $3.3\times10^{32}$ years at 90\% C.L.

\par
Finally, we describe the search for \pknu ; \pipi.
The $\pi^0$ and $\pi^+$ from the $K^+$ decay at rest 
have equal and opposite momenta of approximately 205 MeV/$c$. 
The decay $\gamma$'s from the $\pi^0$ 
reconstruct this momentum.  
The $\pi^+$, barely over Cherenkov threshold with $\beta \approx .86$, 
emits very little Cherenkov radiation.  However it does decay into a muon
 ($\pi^+ \rightarrow \mu^+ \nu_{\mu}$) which decays into a
 positron ($\mu^+ \rightarrow e^+ \nu_e \bar{\nu}_{\mu}$).  Detection of this
 positron is possible.  Furthermore, a small amount of Cherenkov 
radiation from the $\pi^+$, sometimes visible as a collapsed Cherenkov
ring, can be detected in the direction opposite 
that of the $\pi^0$.  To quantify this, we defined 
``backwards charge'' ($Q_b$) as the sum of the photoelectrons
detected by the PMTs which lay within a $40^\circ$
 cone whose axis was the opposite direction of the reconstructed direction of
 the $\pi^0$.  This charge was corrected for light attenuation in the water,
angular dependence of photon acceptance, and photocathode coverage.

\par
To determine the signal region and estimate the detection efficiency, 
a total of 1000 
\pknu ; \pipi MC events were generated, 771 of which had vertices
which reconstructed inside the fiducial volume of the detector.  
For these 205 MeV/$c$ $\pi^0$ events, 
the resolution of vertex fitting was 29 cm and 66\% of the events were
identified as 2-ring events.

 The selection criteria for this type of event were defined:
B1) 2 $e$-like rings,
B2) with 1 decay electron, 
B3) $85$ MeV/$c^2$ $< \mbox{mass}_{\gamma\gamma} < 185$ MeV/$c^2$,
B4) $175$ MeV/$c^2$ $< \mbox{momentum}_{\gamma\gamma} < 250$ MeV/$c^2$ and
B5) $40$ p.e. $<  Q_b < 100$ p.e.
Criteria B1,B3, and B4 required the $\pi^0$ with the monochromatic
 momentum expected.  Criterion B2 required the decay of the
 $\pi^+$ into muon into positron.  Criterion B5 required the
Cherenkov light from the $\pi^+$.
The $\pi^0$ mass resolution was determined to be 135$\pm$21 MeV/$c^2$.
By passing the proton 
decay MC events through these selection criteria, the detection 
efficiency was determined to be 31\%.  The largest contribution to the 
inefficiency was the inefficiency of detection of two $\gamma$-rays from
the decay of the 
$\pi^0$.  Including the kaon branching ratio of $21.2\%$ into $\pi^+\pi^0$,  
the total detection efficiency for this mode was estimated to be $6.5\%$.

\par
Charged current interactions such as
$\nu_{\mu} N \rightarrow \mu N^\prime \pi^0$ 
from atmospheric neutrinos can imitate a kaon decay mode of this type.  
The selection criteria were applied to the sample of atmospheric neutrino
MC. The flux was normalized in the same manner as in the prompt $\gamma$-ray
search.
 The number of background events expected was estimated to 
be 0.7 events/33kt$\cdot$yr. 
Figures \ref{backqvsp}a and b show $|\vec{p}_{\pi^0}|$ vs. backwards charge 
(Q$_b$) for proton decay MC and  atmospheric neutrino MC, respectively.

\par
The selection criteria for this decay mode were applied to the data.
Figure  \ref{backqvsp}c shows the results of the final two cuts.  No
events passed.  The single event which lay close to the cuts was examined 
visually 
with an event display.  We found no additional evidence that it could be a
signal event that fell outside of the cuts.  The backwards charge 
appeared to be a fragment of one of the rings and not from a small 
collapsed ring.  Based on these numbers, the lower limit of the 
partial lifetime of \pipi was estimated to be 
$3.1 \times 10^{32}$ yr at the 90\% C.L. based on Eqn. \ref{lifelim}.

\par
The combined 90\% C.L. upper limit for the number of proton decay
candidates ($x_{limit}$) was calculated by integrating the likelihood 
function to the 90\% probability level:
\begin{equation}
  \frac{ \int_0^{x_{limit}}[\prod_{i=1}^{3}P(N^{obs}_i,N_i(x))]dx}
       { \int_0^{\infty}[\prod_{i=1}^{3}P(N^{obs}_i,N_i(x))]dx}
  =  0.90,\label{lifelimit_eqn}
\end{equation}
\begin{equation}
  N_i(x)=
    X^{BG}_i+
    \frac{\epsilon B_m^i}
         {\sum_{j=1}^{n}[\epsilon_i B_m^j]}x,
\end{equation}
where $P(N,x)$ is the
probability function of Poisson statistics, $N^{obs}_i$ is the number
of observed candidates, $X^{BG}_i$ is the number of estimated background
events, $\epsilon_i$ is the detection efficiency, and $B_{m}^i$ is the
meson branching ratio. The index $i$ stands for the $i$-th method.
The lower limit of partial lifetime was calculated with
\begin{eqnarray}
  \tau/B=\frac{1}{x_{limit}}(\sum_{i=1}^{3}\epsilon_i B_{m}^i)\Lambda,
\end{eqnarray}
where $\Lambda$ is the exposure in proton$\cdot$years.
The combined lower limit of the partial lifetime for
$\tau/B(p \rightarrow \bar{\nu} K^+)$ using the three
independent methods was $7.3 \times 10^{32}$ yr at the 90\% C.L.

\par
The main sources of systematic errors in the calculation of the 
lifetime limits were:
{1) uncertainty in the energy calibration,}
{2) uncertainty in the detection efficiency of the decay electron 
from the stopped muon,}
{3) uncertainty in the atmospheric neutrino fluxes and interaction 
cross sections used in the Monte Carlo simulation,}
{4) uncertainty in the emission probability of prompt $\gamma$-ray from
$^{16}$O.}

For the \pipi mode search, the detection efficiency changed by less than 
$\pm 1\%$ if the reconstructed momentum criterion was shifted by $\pm2.5\%$. 
For the $K^+ \rightarrow \mu^+ \nu_\mu$ mode,
if the reconstructed momentum was shifted by 
$\pm 2.5\%$, the lifetime limit obtained from this analysis was changed 
by $\pm10\%$ at most.
The systematic error of the detection
efficiency of decay electrons from muons was estimated to be 
1.5\%\cite{atmpd_paper}
by comparing the fraction of cosmic-ray muon events with decay electrons 
between MC simulation and real data. 
The uncertainty of the estimation of number of background events came from 
the uncertainty of the neutrino flux and the interaction cross
sections. This uncertainty was estimated by comparing the number of events of 
the normalized atmospheric neutrino background with the real data 
at each reduction step.  
For the prompt $\gamma$-ray tagging method and \pipi mode search,
the data and atmospheric $\nu$ Monte Carlo agreed 
within statistical  errors at each step.  In searching for the 
excess of 236 MeV/$c$ $\mu^+$s, the shape of the momentum 
distribution and the absolute 
normalization was left as a free parameter in the fitting. 
Therefore, the uncertainties of neutrino flux and the interaction
cross-sections did not significantly affect the lower limit 
of the partial lifetime.
The systematic error in the 6.3 MeV $\gamma$-ray emission probability was
estimated to be $\pm 15\%$\cite{GAMSYSERR}. This error directly affected
the detection efficiency of the prompt $\gamma$-ray tagging method.
Also it is expected that there are some other promt $\gamma$-rays,
whose energies are 9.93 MeV, 7.03 MeV, 7.01 MeV\cite{GAMEMIT}.
Including these deexcitation modes, the detection efficiency
of the prompt $\gamma$-ray tagging method should be higher.
However, these emission probabilities are more than 10 times 
smaller than the probability of that of 6.3 MeV $\gamma$-ray and
they have much larger uncertainties in the emission probability,
which are estimated to be larger than 30\%\cite{GAMSYSERR}.
Therefore, we neglected these modes in the analysis. However, these
modes were taken into account in the systematic uncertainty of the
lifetime limit, which was estimated to be $^{+7}_{-3}\%$.

Considering all of these effects, the systematic error of the
lower limit of the partial lifetime was estimated to be $^{+10}_{-8}\%$.
As a result, the lower limit of the partial lifetime for $p \rightarrow 
\bar{\nu}K^+$ was estimated to be $6.7\times10^{32}$yr(90\%C.L.).

In this paper, we have reported the result of a search for proton 
decay into $\bar{\nu} K^+$ in a 33 kt$\cdot$year exposure of
the \SuperK detector. The data are consistent with the background 
expected from atmospheric neutrinos and no evidence for proton decay
was observed. 
We set the lower limit of partial lifetime for
$p \rightarrow \bar{\nu} K^+$ to be $6.7 \times 10^{32}$ yr at 90\% C.L.
This limit is more than six times longer than the previously published
best limit
(1.0 $\times 10^{32}$ yr.(90\% C.L.))\cite{KAM} and will help to constrain
SUSY GUT models.

 We gratefully acknowledge the cooperation of the Kamioka Mining and 
Smelting Company. The Super-Kamiokande experiment was built from and 
has been operated with, funding by the Japanese Ministry of Education, 
Science, Sports and Culture, and the United States Department of Energy.

\begin{figure}[hptb]
\begin{center}
\psfig{file=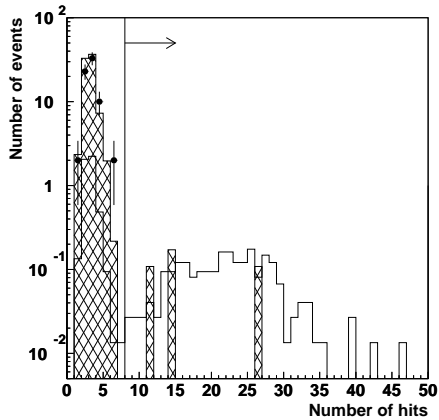,width=6cm}
\caption{Number of hit PMTs within the timing region of the prompt
$\gamma$-ray search in
\mbox{$p \rightarrow \bar\nu K^+, K^+\rightarrow\mu^+ \nu_\mu$}.
Histogram shows the proton decay
Monte Carlo with 6.3MeV $\gamma$-ray, shaded histogram shows 
33kt$\cdot$yr equivalent atmospheric neutrino Monte Carlo events,
and data points with error bars show 33kt$\cdot$yr 
data of Super--Kamiokande. For the proton decay
Monte Carlo, $\tau/B(p \rightarrow \bar{\nu} K^+)$ of $2.1 \times 10^{32}$ yr
is assumed.}
\label{ivpekrk}
\end{center}
\end{figure}

\begin{figure}[hptb]
\begin{center}
\psfig{file=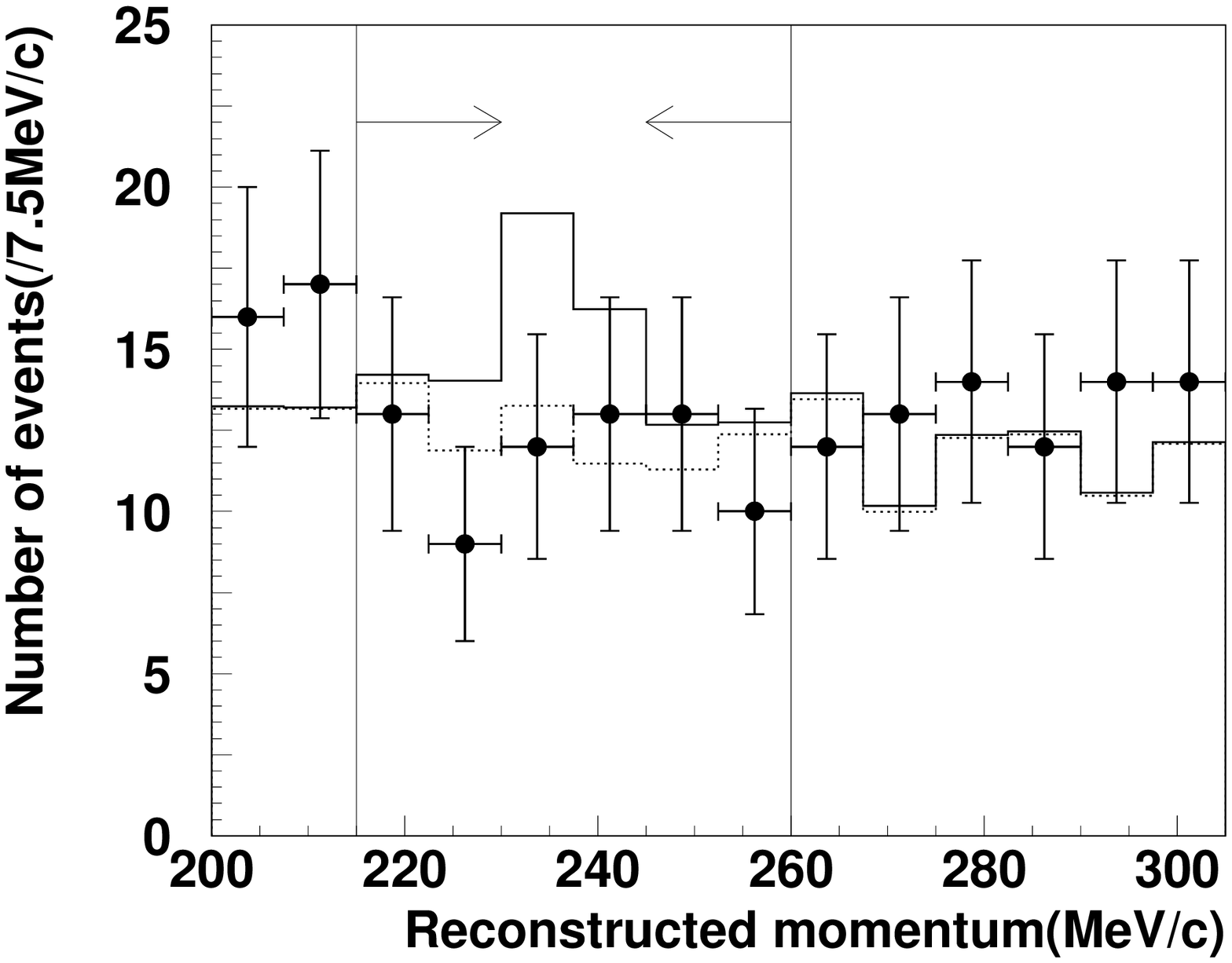,width=6cm}
\caption{Reconstructed momentum distribution. 
Solid (dotted) line shows the estimated 90\% C.L. number of 
{proton decay ($\tau/B(p \rightarrow \bar{\nu} K^+) = 3.3 \times 10^{32}$ yr)
 + atmospheric $\nu$} (atmospheric $\nu$) events;
the black points with error bars show the data with the statistical errors.}
\label{mom_dist_fitresult}
\end{center}
\end{figure}

\begin{figure}
\begin{center}
\psfig{file=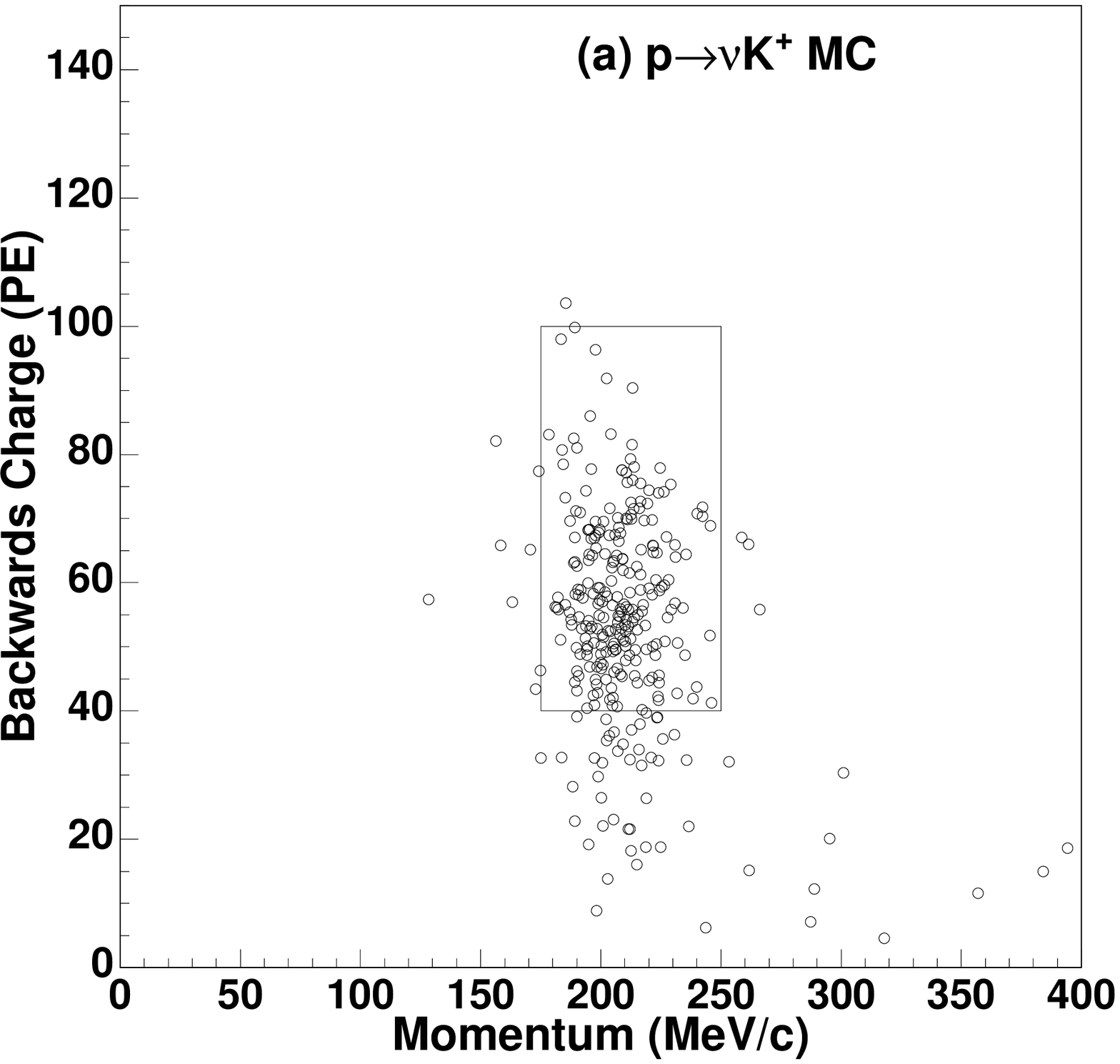,width=6cm}
\psfig{file=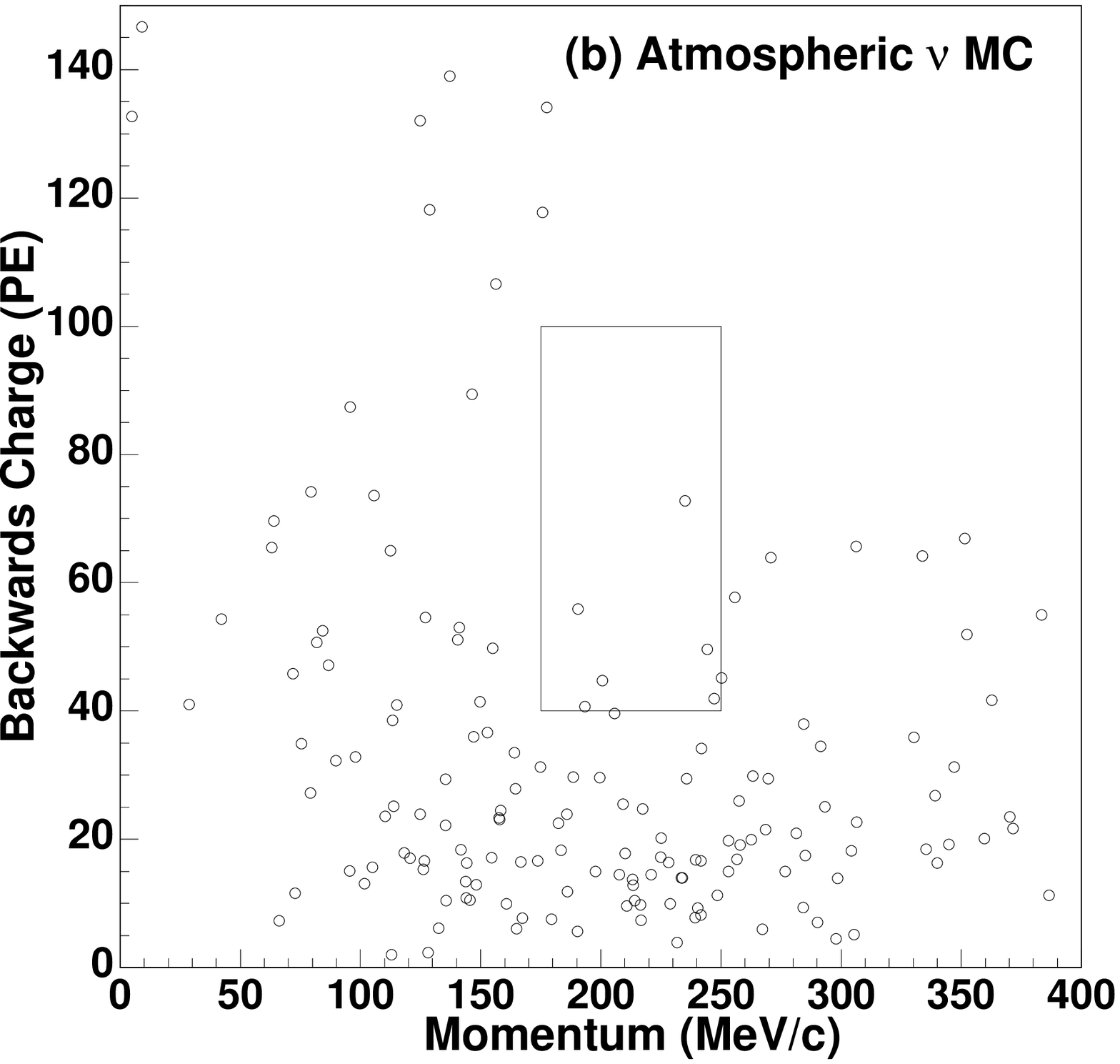,width=6cm}
\psfig{file=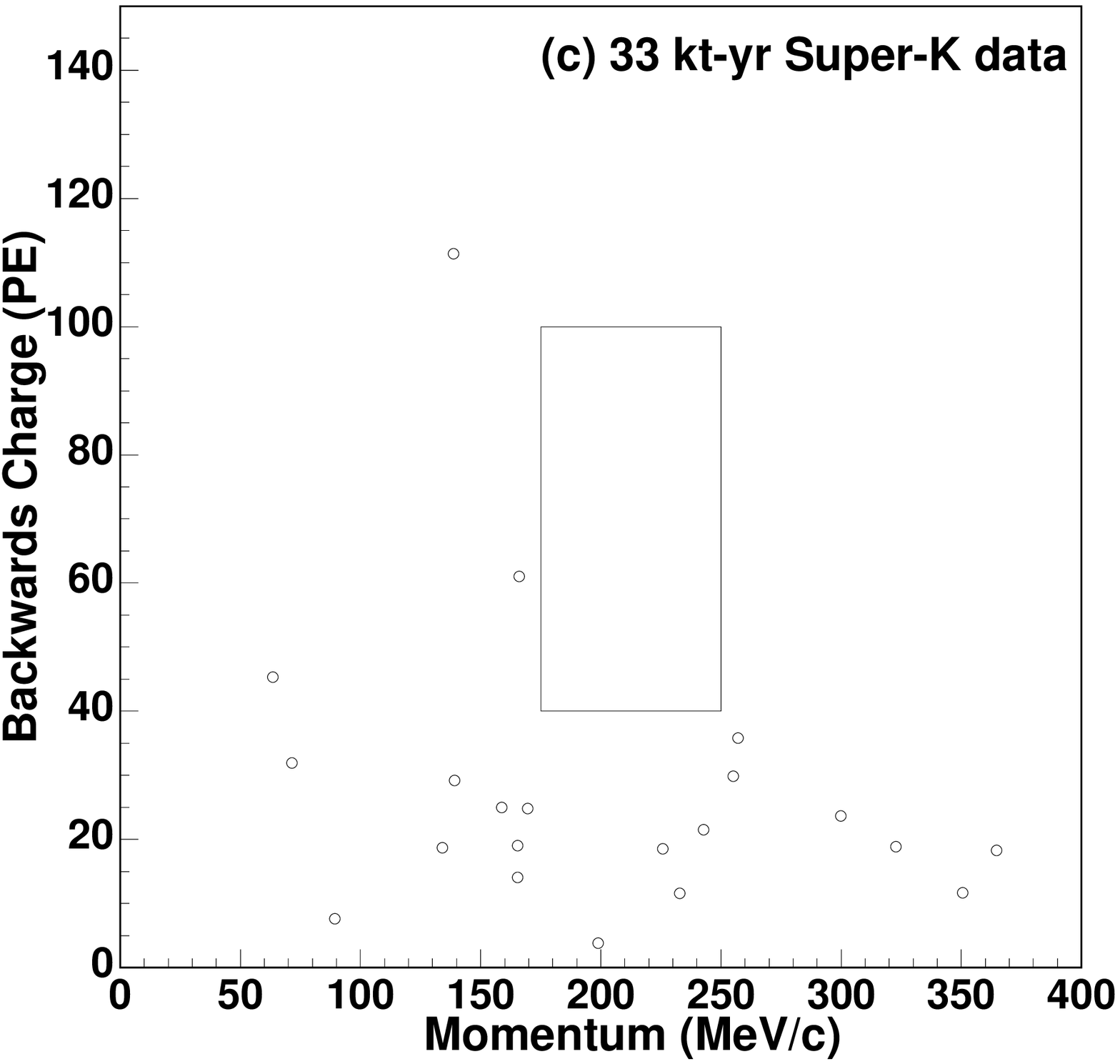,width=6cm}
\caption{\label{backqvsp}Backwards charge versus $\pi^0$ momentum
for (a) $p \rightarrow \bar{\nu} K^+ \mbox{;} K^+
 \rightarrow \pi^+\pi^0$ Monte Carlo,  
(b) 225 kt$\cdot$yr equivalent atmospheric $\nu$ Monte Carlo, and 
(c) 33 kt$\cdot$yr data from Super--Kamiokande.} 
\end{center}
\end{figure}

\end{document}